\documentclass[preprint]{jpsj2}

\def\runauthor{T. Nagao and R. Shiina}

\title{
Studies on X-ray Thomson Scattering from Antiferroquadrupolar
Order in TmTe
}

\author{Tatsuya {\sc Nagao} and Ryousuke {\sc Shiina}$^{1}$}

\inst{Faculty of Engineering, Gunma University, Kiryu 376-8515, Japan\\
$^{1}$Institute of Physics, Kanagawa University, Yokohama 221-8686, Japan
}

\recdate{\today}

\abst{
We study Thomson scattering from
the antiferroquadrupole ordering phase in TmTe.
On the basis of the group theoretical treatment,
we classify the selection rules of the scattering intensity
governed by the orientation of the scattering vector $\textbf{G}$.
Then, numerical verification is performed by invoking the ground states
which are deduced from a $J=\frac{7}{2}$ multiplet model.
The obtained intensity varies drastically depending on the magnitude and
direction of $\textbf{G}$.
We also calculate the scattering intensities under the applied
field for $\textbf{H} \parallel (001)$ and $(110)$.
Their results behave differently when the orientation of
$\textbf{G}$ is changed, which is ascribed to
the difference of their primary order parameters;
$O_{2}^{0}$ and $O_{2}^{2}$ for $\textbf{H} \parallel (001)$ and $(110)$,
respectively.
We make critical comparisons between our results for
TmTe and the experimental ones for CeB$_6$.
First, we assert that the intensities expected from TmTe at several
forbidden Bragg spots
are sufficient enough to be experimentally detected.
Second, their intensities at
$\left(\frac{7}{2}\frac{1}{2}\frac{1}{2}\right)$
differ significantly and may be
attributed to the difference of the order parameters
between the $\Gamma_3$-type ($O_{2}^{2}$ and $O_{2}^{0}$)
and $\Gamma_5$-type ($O_{yz}, O_{zx}$, and $O_{xy}$)
components, respectively.
}

\kword
{Thomson scattering, non-resonant X-ray scattering,
multipole, antiferroquadrupole order, forbidden Bragg spot, TmTe}

\begin{document}
\maketitle

\newcommand{\bJ}{\mbox{\boldmath{$J$} } }
\newcommand{\bI}{\mbox{\boldmath{$I$} } }
\newcommand{\bO}{\mbox{\boldmath{$O$} } }
\newcommand{\bT}{\mbox{\boldmath{$T$} } }
\newcommand{\bH}{\mbox{\boldmath{$H$} } }
\newcommand{\btau}{\mbox{\boldmath{\tau} } }

\newcommand{\ri}{ \rm i }
\newcommand{\rs}{ \rm s }
\newcommand{\rM}{ \rm M }
\newcommand{\rQ}{ \rm Q }
\newcommand{\rN}{ \rm N }
\newcommand{\rZ}{ \rm Z }
\newcommand{\rC}{ \rm C }
\newcommand{\rJ}{ \rm J }
\newcommand{\rT}{ \rm T }

\section{Introduction}
The interplay of orbital and spin degrees of freedom in localized magnetic
materials brings about a wide variety of interesting phenomena.
In many $f$-electron systems, due to the strong coupling between the
spin and
orbital angular momenta, the states are described by the multiplets
of the total angular momentum $J$.
When the symmetry exhibited by the system is sufficiently high, the
multiplet
enables the higher rank multipoles as well as the dipole (rank one) be
active.
In fact, various experimental and theoretical studies have been devoted to
clarify the nature of the ordered phase of multipole order parameters
with rank higher than one.\cite{Santini2009,Kuramoto2009}
Among the most investigated systems,
the materialization of the antiferroquadrupole (AFQ) ordering phase
has been established in the materials such as CeB$_6$ and DyB$_2$C$_2$.

Among many experimental probes, scattering experiments
such as resonant X-ray scattering (RXS) and (non-resonant X-ray) Thomson
scattering
provide very powerful tools to reveal the natures of the higher rank
multipolar
ordering phase.
For example, the AFQ ordering phases in CeB$_6$ and DyB$_2$C$_2$ are
investigated in detail
by means of
RXS\cite{Nakao2001,Yakhou2001,Tanaka1999,Hirota2000,Matsumura2002}
and Thomson
scattering.\cite{Yakhou2001,Adachi2002,Tanaka2005,Staub2006}

On the other hand, the situation of TmTe may still be
rudimentary. This material is believed to show the AFQ order
below $T_{\textrm{Q}}=1.8$ K.\cite{Matsumura1998}
Although many evidences for the AFQ order were gathered in terms
of various experimental probes,\cite{Clementyev1997,Link1998,Mignot2002}
there remain some important issues unsettled yet.
For instance, the crystal electric field (CEF) level scheme,
the component of the primary order parameter,
the nature of the multipolar interaction, and so on.
In order to address such issues,
the approaches in terms of the scattering experiments
may be helpful.
In our previous work, we have determined the CEF
level scheme as $\Gamma_8-\Gamma_6-\Gamma_7$
and analyzed some properties expected from the azimuthal
angle dependence of the RXS intensity, which are useful to distinguish the
type of the order parameter.\cite{Shiina2008}

In this paper, we carry out some investigations on
Thomson scattering expected from the AFQ phase in TmTe.
After introducing the theoretical framework to calculate the
scattering intensity,
we first proceed to classify the selection rules of intensity
governed by the direction of the scattering vector.
In the absence of the applied field, such selection rules as well as
the domain consideration determine the whole intensity.
The intensity exhibits the strong dependence on the magnitude
and orientation of the scattering vector.
We verify the qualitative results with the numerical calculation
performed on the theoretical model developed
in our previous paper.\cite{Shiina2008}
We also investigate how the application of the external field
alters the scattering intensity.
When the field is applied along $(001)$ and $(110)$, the primary
order parameters derived from the model are
$O_{2}^{0}=1/2(2J_{z}^{2}-J_{x}^{2}-J_{y}^{2})$ and
$O_{2}^{2}=\sqrt{3}/2(J_{x}^{2}-J_{y}^{2})$, respectively.
We find their intensities show different behaviors as a function
of the orientation of the scattering vector, reflecting the
difference of their primary order parameters.

We also try to compare the present results with those
obtained for CeB$_6$.\cite{Yakhou2001,Nagao2003}
Although both TmTe and CeB$_6$ exhibit the AFQ ordering phases, it is said
the components of the order parameters are different from each other;
the $\Gamma_3$-type ($O_{2}^{2}$ and $O_{2}^{0}$) in the former and
the $\Gamma_5$-type ($O_{yz}, O_{zx}$, and $O_{xy}$) in the latter.
Here, $O_{yz}=\sqrt{3}/{2}(J_y J_z + J_z J_y), O_{zx}=\sqrt{3}/2(J_z J_x
+ J_x J_z)$, and $O_{xy}=\sqrt{3}/2(J_x J_y + J_y J_x)$.
Our investigation tells that: First, there is a realistic chance to
experimentally detect the Thomson scattering signals in TmTe.
Second, the intensities show
different tendency in both materials at several forbidden Bragg spots,
which may be attributed to difference of the component of the order
parameters.

This paper is organized as follows.
Section 2 is spent to introduce a theoretical framework
in order to calculate the Thomson scattering intensity.
In \S 3, we briefly summarize the CEF scheme concluded from our
previous paper and explain the ground state both in the absence and
presence of the applied magnetic field.
In \S 4, we derive some properties of the Thomson scattering intensities
from the AFQ phase in TmTe,
for instance, its dependence on the direction of the scattering vector
and applied
magnetic field. A comparison of the present results and those obtained
for CeB$_6$
is also found. Finally, \S 5 is devoted to concluding remarks.
Note that a very early stage of the present work is
published elsewhere.\cite{Nagao2009}

\section{Scattering Amplitude of Thomson Scattering\label{sect.2}}
The cross section of Thomson scattering is defined as
\begin{equation}
\left(\frac{d \sigma}{d \Omega} \right) = \left| r_0
(\mbox{\boldmath{$\epsilon$}} \cdot \mbox{\boldmath{$\epsilon$}} ')
f(\textbf{G}) \right|^2,
\label{eq.cross-section}
\end{equation}
where $r_0$ is the classical electron radius.
The directions of polarization for the incident and scattered photons are
denoted by $\mbox{\boldmath{$\epsilon$}}$ and
$\mbox{\boldmath{$\epsilon$}}'$,
respectively.
The inner product
$\mbox{\boldmath{$\epsilon$}} \cdot \mbox{\boldmath{$\epsilon$}} '$
gives non-zero value only when the photon polarization is unrotated;
being unity in the $\sigma-\sigma'$ channel while $\cos(2
\theta_{\textrm{B}})$
in the $\pi-\pi'$ channel where $\theta_{\textrm{B}}$ is the Bragg angle.
The scattering amplitude is described as $f(\textbf{G})$ where
the scattering vector is defined as $\textbf{G}= \textbf{k}'-\textbf{k}$
with $\textbf{k}$ and $\textbf{k}'$ being the wave vectors of the incident
and scattered photons, respectively.

We consider the localized electron system with the $(4f)^N$
configuration. The scattering amplitude may be given by a sum of the
contributions from the localized $4f$ electrons:
\begin{eqnarray}
f(\textbf{G}) &=& \frac{1}{\sqrt{N_0}}
\sum_{j} \sum_{n=1}^{N} \sum_{\mu} p_{\mu}(j)
\textrm{e}^{-i \textbf{G} \cdot \textbf{R}_j} \nonumber \\
& \times&
\langle 0_{\mu}(j) | \textrm{e}^{-i \textbf{G} \cdot \textbf{r}_n }
| 0_{\mu}(j) \rangle,
\label{eq.amplitude.def}
\end{eqnarray}
where $N_0$ is the number of Tm ion sites.
Electron position $\textbf{r}_n$ is measured in the coordinate system
centered at each Tm site $j$.
The $| 0_{\mu}(j) \rangle$ refers to the ground state of the $N$
electrons with
probability $p_{\mu}(j)$ where $\mu$ distinguishes possible degeneracies.
We proceed to rewrite the expectation value part in eq.
(\ref{eq.amplitude.def}),
hence we omit the labels $j$ and $\mu$ in the following.

The numerical evaluation of the amplitude can be easily performed
by utilizing the so-called Rayleigh expansion of the
exponential\cite{Messiah1964}
\begin{equation}
\textrm{e}^{-i \textbf{G} \cdot \textbf{r} }
= 4 \pi \sum_{k=0}^{\infty} (-i)^k j_k(G r) \sum_{k_z=-k}^{k}
Y_{k,k_z}(\Omega) Y_{k,k_z}^{\star}(\Omega_G),
\label{eq.Rayleigh}
\end{equation}
where $j_k$ means the $k$-th order spherical Bessel function
and $G r = |\textbf{G}| | \textbf{r}|$.
The solid angles of $\textbf{r}$ and $\textbf{G}$ are represented
as $\Omega$ and $\Omega_G$, respectively.
Note that the similar treatments are found in the literatures
analyzing Thomson scattering of X-rays from the ordering phase
in CeB$_6$ for $f^{1}$-configuration.\cite{Lovesey2002,Nagao2003,Kono2004}

In expanding the ground state, we employ the total angular momentum basis
involving the radial part, $|J, J_z \rangle$, as follows:
\begin{equation}
| 0 \rangle = \sum_{J_z=-J}^J a(J_z) | J_z \rangle.
\label{eq.state.J}
\end{equation}
where we denote $\left|J, J_z \right\rangle$ as $|J_z \rangle$.
The state $|J_z \rangle$ can be expanded by means of the Slater
determinant constructed by the one-electron spin orbitals for $N$
electrons.
Generally, the evaluation of eq. (\ref{eq.amplitude.def})
from the $N$-electron Slater determinant is tedious,\cite{Keating1969}
and one can employ the formalism
on the basis of the Stevens operator equivalence method.\cite{Amara1998}
When $N=1$ and $13$, however, the situations are quite simple
and we can carry out the evaluation easily.
Reflecting the fact that Tm$^{2+}$ ion is in the $f^{13}$-configuration,
we restrict $N=13$ in the following.

Then, the Slater determinant for thirteen electrons is specified by
the quantum numbers, orbital ($\ell_z$) and spin ($s_z$) angular momenta,
for a single hole which is
the lone unoccupied one-electron spin orbital in each determinant.
Hence the state $|J_z \rangle$ is written in the form of
\begin{equation}
| J_z \rangle = \sum_{\ell_z=-\ell}^{\ell}
\sum_{s_z=-s}^{s} C(J, J_z:\ell, -\ell_z;s,-s_z )
|\ell \ell_z, s s_z \rangle,
\label{eq.state.LS}
\end{equation}
where $C(JJ_z:\ell\ell_z,ss_z)$ is the Clebsch-Gordan (CG) coefficient
with $\ell=3$ and $s=\frac{1}{2}$ for an $f$ electron.
The ket $|\ell\ell_z,s s_z \rangle$ stands for the
Slater determinant for thirteen electrons labeled by the hole quantum
numbers $\ell_z$ and $s_z$. Note that when the ket means one-electron
spin orbital,
the minus signs in the CG coefficient of eq. (\ref{eq.state.LS})
disappear.
The one-electron spin orbital
is described by the product of radial part $R_{4f}(r)$, angular part
$Y_{\ell,\ell_z}$, and spin part $\chi_{s,s_z}$.
By combining this and eq. (\ref{eq.state.J}) with eq. (\ref{eq.state.LS}),
we can continue the evaluation
of the expectation value of eq. (\ref{eq.Rayleigh}), the detail of which
is relegated to Appendix.
The result is summarized as
\begin{equation}
\sum_{n=1}^{13}
\langle 0 | \textrm{e}^{-i\textbf{G} \cdot \textbf{r}_n} | 0 \rangle
= 14 \langle j_0(G) \rangle - \sum_{J_z, J_z'}
a^{\star}(J_z) a(J_z') f_{J_z, J_z'}.
\label{eq.exponent.1}
\end{equation}
Here, we have introduced the amplitude matrix as
\begin{eqnarray}
f_{J_z,J_z'} &\equiv&
\sum_{\ell_z, \ell_z', s_z}
C(J J_z:\ell \ell_z, s s_z) C(J J_z':\ell \ell_z', s s_z) \nonumber \\
&\times& \sqrt{4 \pi} \sum_{k=0}^{\infty}
(-i)^k \sqrt{2 k + 1} \langle j_k(G) \rangle
Y_{k,\ell_z-\ell_z'}(\Omega_G) \nonumber \\
&\times& c^{k}(\ell \ell_z, \ell \ell_z'),
\label{eq.fjj}
\end{eqnarray}
where the Gaunt coefficient is defined by
\begin{equation}
c^k(\ell \ell_z, \ell \ell_z') =
\sqrt{\frac{4\pi}{2k+1}} \int
Y_{\ell,\ell_z}^{\star}(\Omega) Y_{k,\ell_z-\ell_z'}(\Omega)
Y_{\ell, \ell_z'} d \Omega, \label{eq.Gaunt}
\end{equation}
and
\begin{equation}
\langle j_k(G) \rangle = \int_0^{\infty}
r^{2} j_k(Gr) R_{4f}^2(r) dr.
\label{eq.jGr}
\end{equation}
Since the first term in eq. (\ref{eq.exponent.1}) is independent of the
ground state, it has no contribution to the scattering
intensity as far as $\textbf{G}$ is chosen as antiferro-type spot.

The Gaunt coefficients are evaluated by means of the Wigner $3j$ symbols.
In the present case of $f$ electron system with $\ell$ being fixed to
three,
only the terms for $k=0,2,4$, and $6$ are relevant.
As a consequence, $f_{J_z,J_z'}$ and eventually the scattering
amplitude itself are invariant under the
transformation $\textbf{G} \leftrightarrow
-\textbf{G}$. In this context, we do not discriminate
between $\textbf{G}$ and $-\textbf{G}$ in the present work,
in particular,
when we perform numerical evaluation of the scattering intensities.
Notice that we can verify the symmetry relations exhibited by the
amplitude matrix
elements as follows
\begin{equation}
f_{J_z,J_z'} = f_{J_z',J_z}^{\star} =(-)^{J_z-J_z'} f_{-J_z,-J_z'}^{\star}
=(-)^{J_z-J_z'} f_{-J_z',-J_z}.
\label{eq.fJJ.relation}
\end{equation}
A remaining task is to calculate the coefficient $a(J_z)$'s.
We briefly summarize the methods and results in the next section.

\begin{figure}[t]
\begin{center}
\includegraphics[width=8cm]{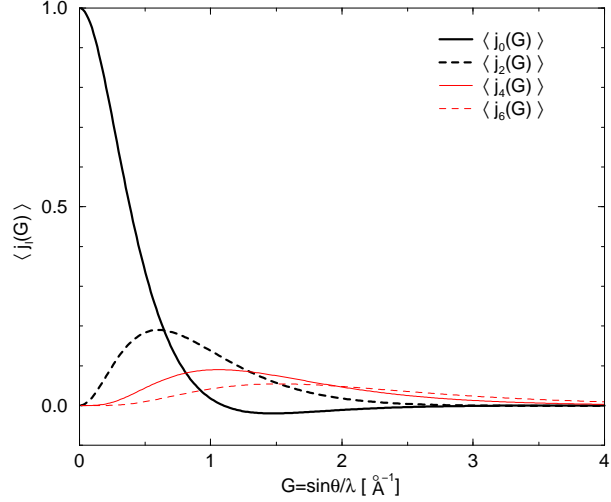}
\end{center}
\vspace{0.0cm}
\caption{
(Color online) Radial integrations of spherical Bessel
function $\langle j_k(G) \rangle$ for divalent Tm ion
as a function of $G=|\textbf{G}|$.
The bold solid (black), bold dotted (black), thin solid (red),
and thin dotted (red) lines for $k=0, 2, 4$, and $6$, respectively.
}
\label{fig.1}
\end{figure}

\section{Ground State\label{sect.3}}
Thulium telluride (TmTe) is a magnetic semiconductor crystallized in a
cubic,
NaCl structure with a lattice constant of $a =6.35 \textrm{\AA}$.
The Tm ion is in a divalent state with one $4f$ hole [$(4f)^{13}$]
configuration
($ ^{2}F_{\frac{7}{2}}$).
The radial part of the wave function $R_{4f}(r)$ we use
is calculated within the Hartree-Fock approximation
for Tm$^{2+}$.\cite{Cowan1981}
The $\langle j_k(G)\rangle$'s are evaluated by means of $R_{4f}(r)$
as shown in Fig. \ref{fig.1}.
Then, the angular part of the wave functions are prepared as follows.

Under the cubic CEF potential, the ground multiplet
spanned by $J=\frac{7}{2}$ subspace is split into
two doublets $\Gamma_6$ and $\Gamma_7$ and a quartet $\Gamma_8$.
Since their total separation is believed to be around 15
K,\cite{Clementyev1997}
we should retain all the bases.
In the previous paper, we have introduced a model Hamiltonian
on the $J=7/2$ multiplet basis to describe the phase diagram
for TmTe.\cite{Shiina2008} Analyzing carefully an interplay among
the CEF potential, the Zeeman energy, and multipolar interactions,
we have concluded that the $\Gamma_3$ AFQ order parameter and
a CEF level structure $\Gamma_8$-$\Gamma_6$-$\Gamma_7$ naturally
explain the observed field dependence and anisotropy of the phase
diagram. In the present analysis, the intensity of X-ray scattering
is calculated by using the mean-field ground state derived from the
same Hamiltonian. Here, let us summarize the Hamiltonian
and its mean-field results briefly.

The basic assumption in the model is that the original {\it fcc}
lattice of Tm ions can be decoupled to four distinct {\it sc}
sublattices.\cite{Shiina1999.1}
Then, the model on the {\it sc} lattice is defined by a sum
of three parts, $ H = H_{\rC} + H_{\rZ} + H_{\rQ} $, where
\begin{subequations}
\begin{align}
H_{\rC} & = W \sum_i \left[
x \frac{ O_4(i) }{ F_4 } + ( 1 - |x| ) \frac{ O_6(i) }{ F_6 }
\right],
\label{eq.11.a} \\
H_{\rZ} & = - g \mu_B \sum_i \bJ (i) \cdot \bH,
\label{eq.11.b} \\
H_{\rQ} & = D_{\rQ} \sum_{ \langle ij \rangle }
\left[
O_{2}^{0}(i) O_{2}^{0}(j) + O_{2}^{2}(i) O_{2}^{2}(j)
\right]. \label{eq.11.c}
\end{align}
\end{subequations}
$H_{\rC}$ is the CEF Hamiltonian defined in ref. \citen{Lea1962},
and $H_{\rZ}$ is the Zeeman energy in the magnetic field $\bH$
with $g=8/7$ being the Land\'e $g$ factor for $J=7/2$.
The $\Gamma_3$ AFQ interaction relevant for TmTe is given by
$H_{\rQ}$, where the summation over $\langle ij \rangle$ is
restricted to the nearest-neighbor sites in the {\it sc} lattice.
For simplicity we do not consider influences of field-induced
multipoles in this model Hamiltonian.

Concerning the parameters in the CEF Hamiltonian $H_{\rC}$,
we assume $W=-0.417$K and $x=0.5$ which lead to
a level scheme $\Gamma_8$(0K)-$\Gamma_6$(5K)-$\Gamma_7$(10K).
This is nothing but scheme (b) in ref.\citen{Shiina2008},
namely the most promising CEF level scheme for TmTe.
The quadrupole coupling constant is determined so as to
give the transition temperature $T_{\rQ}=4$K at zero field,
for the fixed CEF level scheme. We expect that the transition
temperature must be suppressed and becomes closer to the
real value $T_{\rQ}=1.8$K when the strong fluctuation is
taken into account.

Applying the mean field approximation for the AFQ interaction,
one can determine the stable order parameters depending
on the direction of the magnetic fields.\cite{Shiina2008}
It is shown that the $O_{2}^{0}$ order appears in $\bH || (001)$
whereas the $O_{2}^{2}$ order is stabilized in $\bH || (110)$.
On the other hand, $O_{2}^{0}$ and $O_{2}^{2}$ are almost degenerate
at zero field and in $\bH || (111)$ due to high symmetry.
Thereby, we assume the $O_{2}^{2}$ order at zero field in this study,
because the observed field-induced antiferromagnetic structure in $\bH
|| (110)$
indicating $O_{2}^{2}$ are continuously connected to the zero
field.\cite{Link1998} These mean field analyses obviously
provide the ground state wave function at each sublattice site,
which can be used to calculate the X-ray scattering intensity
at zero temperature.

Finally, we shall briefly comment on the properties of the AFQ
domains within the model described by eqs.
(\ref{eq.11.a}) $\sim$ (\ref{eq.11.c}).
As discussed above, the model leads
to a simple antiferro-type structure characterized by a wave vector
$\textbf{K}_1 = (111)$ in units of $\pi/a$.
Although this wave vector is unique on the $sc$ lattice,
it allows degeneracy on the $fcc$ lattice, with
$\textbf{K}_2=(\overline{1}11)$, $\textbf{K}_3=(1\overline{1}1)$,
and $\textbf{K}_4=(\overline{11}1)$. This degeneracy is equivalent
to the degeneracy when combining four decoupled $sc$ sublattices
to a single $fcc$ lattice. Therefore, the four $K$ domains
remain to be unchanged in the present model even when the magnetic
field is applied. The stability of the $K$ domains is determined
exclusively by a subtle inter-sublattice interaction.\cite{Shiina1999.2}
In the present paper, we will present the results of each $K$ domain
and will not discuss details on the stability problems.

\section{Thomson Scattering Intensities}

\subsection{Remarks on scattering from K domain \label{sect.4.1}}

The scattering vector in Thomson scattering to detect
the antiferro-type ordering pattern in TmTe is simply described
as $\textbf{G}=(2h+1,2k+1,2\ell+1)$ with $h, k$, and $\ell$ being
integers, in units of $\pi/a$. In this case, the phase factor
appeared in eq. (\ref{eq.amplitude.def})
becomes $\textrm{e}^{-i \textbf{G} \cdot \textbf{R}_j} = +1$ or $-1$
depending on which sublattice $\textbf{R}_j$ belongs to.
We can verify that for any antiferro-type $\textbf{G}$, there exists
only one $m$ among $1 \sim 4$ which satisfies
$\textrm{e}^{-i \textbf{G} \cdot \textbf{R}_j} =
\textrm{e}^{-i \textbf{K}_m \cdot \textbf{R}_j}$ at every $\textbf{R}_j$.
Thus, the scattering amplitude remains finite only from
the $K$-domain which satisfies this relation.
In this sense, the scatterings from distinct $K$-domains
should be identified by the corresponding scattering vectors.

Here, we calculate the intensity for the perfect single $K$-domain,
which is picked up by the scattering vector.
This should be kept in mind when we compare the calculated results
with the experimental ones.
If each $K$-domain would have nearly the same population,
our results overestimate factor four.

\subsection{Dependence on the direction of scattering
vector\label{sect.4.2}}

We exploit a group theoretical analysis on the scattering amplitude
$f(\textbf{G})$
[eq. (\ref{eq.amplitude.def})].
The amplitude is invariant under the symmetry operations
keeping both crystal and $\textbf{G}$ unchanged.
Under cubic symmetry ($O_h$), $f(\textbf{G})$ is constructed by the
quantities belonging to the identical ($\Gamma_1$) representation
in a point group $O_h \times \textbf{G}$.
Here, some symmetry operations in $O_h$ are forbidden by assuming
an artificial strain along $\textbf{G}$ in $O_h \times \textbf{G}$.
Since $f(\textbf{G})$ is expanded by a linear combination of
the terms with even rank, only the even rank multipole operators belonging
to the $\Gamma_1$ representation in $O_h \times \textbf{G}$ contribute to
the scattering intensity.
This is a striking difference compared with a starting point of the
similar analysis for the magnetic neutron scattering form factor
where the unprojected scattering operator behaves
as the odd rank multipole operators.\cite{Shiina2007}
In the present case of the antiferro-type ordering phase,
it corresponds to detect the $\Gamma_1$ representation from rank two
operator. This is easily confirmed by re-expressing eq.
(\ref{eq.exponent.1}) as
\begin{eqnarray}
& & \sum_{n=1}^{13}
\langle 0 | \textrm{e}^{-i\textbf{G} \cdot \textbf{r}_{n}} | 0 \rangle
= 13 \langle j_0 (G) \rangle
\nonumber \\
&-& \sum_{k=2,4,6}
\sum_{J_z,J_z '}
a^{\star}(J_z) a(J_z ') B^{k}(J_z, J_z ')
\langle j_{k}(G) \rangle,
\label{eq.exponent.3}
\end{eqnarray}
where coefficient $B^{k}(J_z,J_z')$ is obtained by
eliminating expression $\sum_{k=0}^{\infty} \langle j_k(G)\rangle$
from the right hand side of eq. (\ref{eq.fjj}).
The first term is canceled
by the contributions from two sublattices, which leaves the expansion
starting with the term $k=2$.

Equation (\ref{eq.exponent.3}) indicates the presence of the contributions
from the terms of rank four and six.
However, qualitative behavior of the whole intensity is well understood by
that from the leading term
proportional to $\langle j_2(G) \rangle$.
The reasons are two fold.
First, because the symmetry properties
of the coefficients $B^{2}(J_z,J_z'), B^{4}(J_z,J_z')$, and $B^{6}(J_z,J_z')$ 
deduced from relation
similar to eq. (\ref{eq.fJJ.relation}) are the same one another,
the latter two terms do not 
give rise to qualitatively new properties
which are absent for $B^{2}(J_z,J_z')$ term alone.
It means their influence on the total intensity is quantitative,
not qualitative.
Second, it turns out from the numerical calculations in the following
subsections that
the contribution from the term proportional to $\langle j_2(G) \rangle$
dominates the intensity.
Therefore, though our numerical calculation shall include the
contributions from
the terms with rank four and six, we proceed to make a group theoretical
consideration
deduced only from the rank two term and derive some selection rules
which qualitatively explain the behavior of the whole intensity.

In a point group $O_h$, rank two quadrupole operators
are $\Gamma_3$-type $(O_{2}^{0}, O_{2}^{2})$ and
$\Gamma_5$-type $(O_{yz}, O_{zx}, O_{xy})$.
For $\textbf{G} \parallel (001), (111)$, and $(110)$,
the components to be invariant under $O_h \times \textbf{G}$
are $O_{2}^{0}$, $(O_{yz}+O_{zx}+O_{xy})/\sqrt{3}$, and
$O_{2}^{0}$ and $O_{xy}$, respectively.
Note that this is the same as symmetry lowering of quadrupoles
by the magnetic field, as discussed in ref. \citen{Shiina1997}.
Although antiferro-type $\textbf{G}$ spot does not exist in the
$(001)$ nor $(110)$ directions, they
are interpreted as the limiting cases of
the spots, for example, at
$\textbf{G}=(1,1,2h+1)$ and $(2h+1,2h+1,1)$, respectively.

In the absence of the applied field, the order parameter is $O_{2}^{2}$ as
explained in the previous section.
In the cubic symmetry, two more independent primary order parameters are
obtained by rotating $O_{2}^{2}$ by an angle $\pm 2\pi/3$ 
about the wave vector specifying the $K$-domain.
The domains specified by these primary order parameters
are called as $S$-domains. 
In the $K_1$-domain, for instance, the primary order parameters
of three $S$-domains become $O_2^{2}$, $(\sqrt{3}O_2^{2}-O_2^{0})/2$, and
$-(\sqrt{3}O_2^{2}-O_2^{0})/2$.
Then, for a given $\textbf{G}$,
if the $\Gamma_1$ representation contains $O_{2}^{0}$ and/or $O_{2}^{2}$,
we can expect the Thomson scattering intensity remains finite.
We proceed to our investigation assuming that
three $S$-domains have the equal population in each $K$-domain.
In the following, the numerical results are presented for the
$\sigma-\sigma'$ channel when unspecified.

\begin{figure}[t]
\begin{center}
\includegraphics[width=8cm]{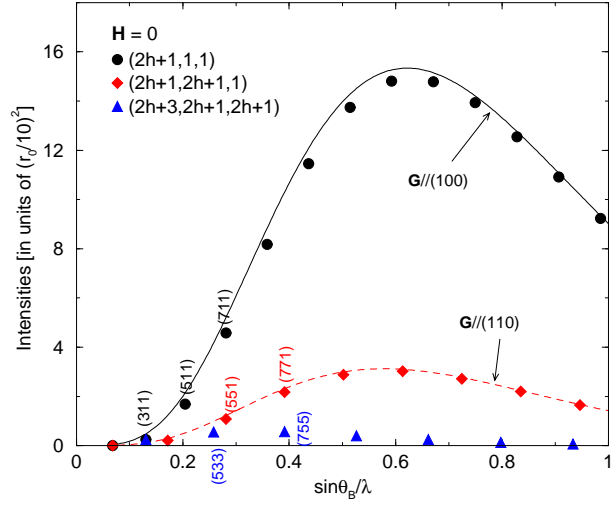}
\end{center}
\vspace{0.0cm}
\caption{
(Color online) The Thomson scattering intensities per Tm ion
under no applied field in the $\sigma-\sigma'$ channel.
The (black) circles, (red) diamonds, and (blue) triangles
are the intensities for $\textbf{G}=(2h+1, 1, 1), (2h+1, 2h+1, 1)$,
and $(2h+3, 2h+1, 2h+1)$, respectively, with $h=0,1,2, \cdots$.
The (black) solid and (red) dotted lines show the limiting curve
for $\textbf{G} \parallel (100)$ and $(110)$, respectively.
}
\label{fig.2}
\end{figure}

Let us consider $\textbf{G}||(111)$ at first. It is clear that
the $\Gamma_1$ representation $(O_{yz}+O_{zx}+O_{xy})/\sqrt{3}$
in $O_h \times \textbf{G}$ is not involved in the order parameter.
Therefore, the scattering is forbidden in this case.
Then, we examine the intensity at $\textbf{G}=(2h+3,2h+1,2h+1)$,
which approaches to $\textbf{G} \parallel (111)$
in the limit of $h \rightarrow \infty$.
As shown in Fig. \ref{fig.2}, the intensity obtained at
$(2h+3,2h+1,2h+1)$ is very tiny as expected.

Second, we consider $\textbf{G}=(001)$.
Two of the three $S$-domains whose order parameters include $O_{2}^{0}$
can give finite intensities in this case. Obviously,
the same result is expected for $\textbf{G}=(100)$ and (010).
In Fig. \ref{fig.2}, we plot the calculated intensity
at $(2h+1,1,1)$, together with the result for continuous
$\textbf{G} \parallel (100)$. The latter is evaluated by assuming
$\textrm{e}^{-i\textbf{G} \cdot \textbf{R}_j}$
in eq. (\ref{eq.amplitude.def})
being $+1$ or $-1$ corresponding to $\textbf{R}_j$'s sublattice.
As seen from Fig. \ref{fig.2}, the limiting curve is in good accordance with
the intensities obtained at the real antiferro-type spots
$\textbf{G}=(2h+1,1,1)$ even when $h$ is small.

Then, for another limiting case, $\textbf{G} \parallel (110)$,
the discussion similar to that for $(100)$ is easily confirmed.
That is, the direction $(110)$
is considered as the limiting direction of an antiferro-type
spot $\textbf{G}=(2h+1,2h+1,1)$. The scattering intensity
at $(2h+1,2h+1,1)$ is equal to those at the
corresponding spots at $(1,2h+1,2h+1)$
and $(2h+1,1,2h+1)$. We verify these results and display
the curve together with that for $\textbf{G} \parallel (110)$
also in Fig. \ref{fig.2}.

\subsection{Dependence on the direction of applied field\label{sect.4.3}}

\begin{figure}[t]
\begin{center}
\includegraphics[width=8cm]{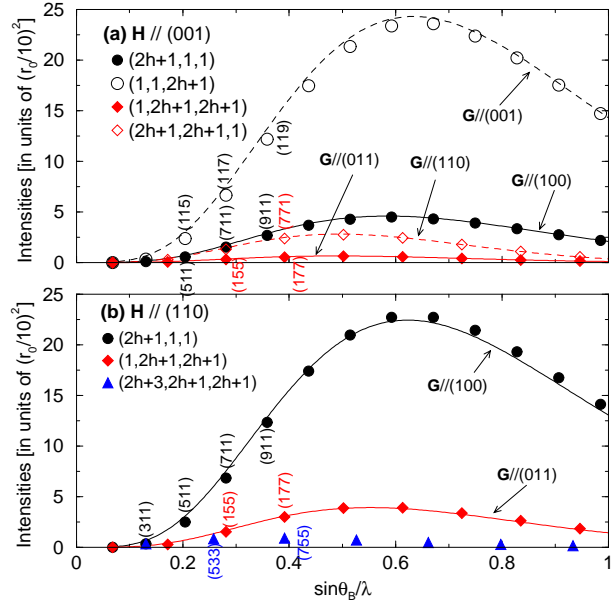}
\end{center}
\vspace{0.0cm}
\caption{
(Color online) The Thomson scattering intensities per Tm ion
under the applied field along (a) $(001)$ and (b) $(110)$ directions
in the $\sigma-\sigma'$ channel.
The (black) filled and open circles are
the intensities for $\textbf{G}=(2h+1, 1, 1)$ and $(1, 1, 2h+1)$,
respectively, with $h=0,1,2, \cdots$.
While the (red) filled and open diamonds, and (blue) filled triangles
are those for
$\textbf{G}= (1,2h+1,2h+1), (2h+1,2h+1,1)$, and
$(2h+3, 2h+1, 2h+1)$, respectively.
The (black) solid and dotted lines show the limiting curves
for $\textbf{G} \parallel (100)$ and $(001)$, respectively.
The (red) solid and dotted lines show the limiting curves
for $\textbf{G} \parallel (011)$ and $(110)$, respectively.
}
\label{fig.3}
\end{figure}

When external field is applied to the system, usually, degeneracies
associated with the $S$-domains are lifted depending on the direction
of the field, which also removes the equivalence of the intensities
under no external field for several high-symmetry $\textbf{G}$ directions.
For instance, for $\textbf{H} \parallel (001)$,
the primary order parameter of the ground state becomes $O_{2}^{0}$.
The identity representation in $O_h \times \textbf{G}$ is
$O_{2}^{0}, (\sqrt{3}O_{2}^{2}-O_{2}^{0})/2$, and $(-\sqrt{3}O_{2}^{2}-O_{2}^{0})/2$
for $\textbf{G} \parallel (001), (100)$, and $(010)$, respectively.
Then, we expect the intensities for $\textbf{G} \parallel (100)$ and
$(010)$ are equivalent and weaker than that for $\textbf{G} \parallel
(001)$.
For corresponding antiferro-type
$\textbf{G}=(1,1,2h+1), (2h+1,1,1)$, and $(1,2h+1,1)$,
the intensities for the latter two are
the same, which are weaker than that for the former one as displayed in
Fig. \ref{fig.3} (a).
Next, we consider the limiting directions
$\textbf{G} \parallel (110), (011)$, and $(101)$.
Since the identity representations belonging to the $\Gamma_3$ block
for them are equivalent
to those for $\textbf{G} \parallel (001), (100)$, and $(010)$,
respectively, the same relations hold.
That is, the intensities for $\textbf{G} \parallel (011)$ and $(101)$
are the same, which are weaker than that for $\textbf{G}\parallel (110)$.
Similarly,
the intensities at $\textbf{G}=(1,2h+1,2h+1)$
and $(2h+1,1,2h+1)$ spots
are the same, which are weaker than the one at $(2h+1,2h+1,1)$ spot.
The intensities at $\textbf{G}=(2h+3,2h+1,2h+1)$
are negligible and we omit them from Fig. \ref{fig.3} (a).

For $\textbf{H} \parallel (110)$,
the primary order parameter of the ground state is $O_{2}^{2}$.
From the analysis for the limiting cases, the series
$\textbf{G} \parallel (100)$ and $(010)$ include $O_{2}^{2}$ as the
$\Gamma_1$ representation while $\textbf{G} \parallel (001)$ does not.
Then,
the intensities at $\textbf{G}=(2h+1,1,1)$
and $(1,2h+1,1)$ are expected to be the same
while tiny intensity, if any, is brought about
at $\textbf{G}=(1,1,2h+1)$.
These tendencies are confirmed numerically as shown in Fig. \ref{fig.3} (b).
Similarly, the intensities at $\textbf{G}=(1,2h+1,2h+1)$
and $(2h+1,1,2h+1)$ give the same values while that at
$\textbf{G}=(2h+1,2h+1,1)$ is essentially zero.

\subsection{Comparison with the CeB$_6$'s results}
Two $f$ electron systems CeB$_6$ and TmTe share some
apparent similarities. First, both materials are cubic systems.
Second, they are considered as the one $f$-particle systems:
In CeB$_6$, Ce$^{3+}$ ion is in the $f^{1}$ configuration in the
electron picture, while
in TmTe, Tm$^{2+}$ ion is in the $\underline{f}^{1}$ configuration
in the hole picture.
Third, they both exhibit AFQ ordering phases below the critical
temperature.
On the basis of these nominal resemblances, our main focus is to
find the differences they may show.
One obvious difference is the $J$ value. Owing to the Hund's rule,
$J=\frac{5}{2}$ and $\frac{7}{2}$ in CeB$_6$ and TmTe, respectively.
Under the cubic circumstances and inferred from the CEF splitting,
their ground states are spanned by
one $\Gamma_8$ quartet in the former, and two doublets ($\Gamma_6$ and
$\Gamma_7$) and one $\Gamma_8$ quartet in the latter.
Those differences may reflect on the
differences of the Thomson scattering amplitude
between TmTe and CeB$_6$.
Due to the difference of the value of $J$ and corresponding
difference of the bases used,
CeB$_6$ does not include the term proportional to $\langle j_6(G) \rangle$
while TmTe does.

Other than this difference, their differences tend to be quantitative ones.
Among them, we examine the experimental results presented by
Yakhou \textit{et al}.
They reported the ratios of the Thomson scattering intensities
at $\textbf{G}=(511)$ and $(711)$ to that at
$\textbf{G}=(533)$ from the AFQ phase in CeB$_6$
under no applied field.\cite{Yakhou2001}
The results are $I_{\pi-\pi'}^{\textrm{exp.}}(511)/
I_{\pi-\pi'}^{\textrm{exp.}}(533) \simeq
I_{\pi-\pi'}^{\textrm{exp.}}(711)/ I_{\pi-\pi'}^{\textrm{exp.}}(533)
\simeq 0.02$
in the $\pi-\pi'$ channel.
Here, the scattering intensity in the $\mu-\mu'$ channel
at scattering vector $\textbf{G}=(2h+1,2k+1,2\ell+1)$
is denoted as $I_{\mu-\mu'}(2h+1,2 k+1,2 \ell+1)$.
With and without the superscript 'exp.' distinguish
between the experimental data
and the theoretical ones, respectively.
The intensities measured in the $\pi-\pi'$ channel
involve the factor $\cos^{2}(2\theta_{\textrm{B}})$, which
depends on the photon energy.
When we compare the experimental data for CeB$_6$
with those obtained from TmTe,
it is convenient to eliminate this factor, which leads to
the ratios expected from the measurement in the $\sigma-\sigma'$ channel.
Then, the experimental data are interpreted as
$I_{\sigma-\sigma'}^{\textrm{exp.}}(511)/
I_{\sigma-\sigma'}^{\textrm{exp.}}(533)\simeq 1.22$ and
$I_{\sigma-\sigma'}^{\textrm{exp.}}(711)/
I_{\sigma-\sigma'}^{\textrm{exp.}}(533)\simeq 0.01$
in the $\sigma-\sigma'$ channel.

To begin with, we comment on
a possibility of the experimental detection of the
Thomson scattering signals from TmTe.
The strongest signal in Yakhou \textit{et al}.'s data
for CeB$_6$ is obtained at $(533)$.\cite{Yakhou2001}
Since they did not present the absolute value of
$I_{\pi-\pi'}^{\textrm{exp.}}(533)$,
we interpret the theoretical intensity at this spot for CeB$_6$
is strong enough to be detected experimentally.
Then, our previous evaluations for CeB$_6$ correspond to
$I_{\sigma-\sigma'}(533)=
1.72 \times 10^{-3}$ from $O_{zx}$ and $O_{xy}$ phases,
while $I_{\sigma-\sigma'}(533)=3.26 \times 10^{-3}$ from $O_{yz}$ phase
in the absence of the external field.\cite{Nagao2003}
These values are measured in units of
$r_0^{2}$ per Ce ion site.
Although we must take the population of the $S$-domains into account
when we compare the theoretical results with the experimental ones,
it may be reasonable to claim the intensity around $1.0 \times 10^{-3}$
is detectable.
In the same units, our present numerical calculation tells that
$I_{\sigma-\sigma'}(533)=5.63 \times 10^{-3}$ for TmTe.
Intensities at another spots such as $(511)$ and $(711)$ give
three to eight times larger than that at $(533)$.
Thus, we assert that the experimental detection of the Thomson
scattering intensity in TmTe is realistically attainable.

Next, we investigate the ratios. 
For CeB$_6$, the magnitudes of the calculated intensities
are compatible with the tendency in the experiments. That is,
$I_{\pi-\pi'}(533)$ is several orders of magnitude stronger 
than $I_{\pi-\pi'}(511)$ and $I_{\pi-\pi'}(711)$. 
Precisely, $I_{\pi-\pi'}(511)/$$I_{\pi-\pi'}(533)$
$\simeq$ $10^{-3}$ and $I_{\pi-\pi'}(711)/$$I_{\pi-\pi'}(533)$
$\simeq$ $10^{-2}$ in our calculation$^{16}$ while 
$I_{\pi-\pi'}^{\textrm{exp}.}(511)/$$I_{\pi-\pi'}^{\textrm{exp}.}(533)$
$\simeq$ 
$I_{\pi-\pi'}^{\textrm{exp}.}(711)/$$I_{\pi-\pi'}^{\textrm{exp}.}(533)$
$\simeq$ $10^{-2}$ in the experiment.$^{4}$
In a qualitative sense, we believe the difference at $\textbf{G}=(511)$
is irrelevant considering the given circumstances such
as a lack of information on the domain population,
the weak signals at $\textbf{G}=(511)$ and $(711)$,
and so on.
Our calculations show
$I_{\sigma-\sigma'}(511)/I_{\sigma-\sigma'}(533) \simeq 3.01$ and
$I_{\sigma-\sigma'}(711)/I_{\sigma-\sigma'}(533) \simeq 8.14$
from the $O_{2}^{2}$ phase in TmTe.\cite{errata}
That is, $I_{\sigma-\sigma'}^{\textrm{exp.}}(711)$ is much smaller than
$I_{\sigma-\sigma'}^{\textrm{exp.}}(511)$ and
$I_{\sigma-\sigma'}^{\textrm{exp.}}(533)$ for
CeB$_6$, while these three quantities have nearly the same magnitudes
for TmTe.
We believe
this difference is easily recognized if the measurements are
available in TmTe.

Note that the difference may be attributed to that of the
nature between the $\Gamma_5$-type order parameters
and the $\Gamma_3$-type order parameters.
Actually, we obtain one corroborating evidence that
the ratios obtained from CeB$_6$ assuming one of $\Gamma_3$-type
order parameters $O_{2}^{2}$ become
$I_{\sigma-\sigma'}(511)/I_{\sigma-\sigma'}(533)\simeq 6.07$ and
$I_{\sigma-\sigma'}(711)/I_{\sigma-\sigma'}(533) \simeq 5.68$,
which are similar to the TmTe's values.
Hence our concern is to understand why $I_{\sigma-\sigma'}(711)$ gives
much larger value in the $\Gamma_3$-type states than
that in the $\Gamma_5$-type states.
To this aim, we invoke the group theoretical consideration
developed in \S \ref{sect.4.2}.
The series $\textbf{G}=(2h+1,1,1)$ approaches to $(100)$
in the limit $h \rightarrow \infty$.
The intensity of the latter is equivalent to that of $(001)$.
Since the scattering vector $\textbf{G} \parallel (001)$ detects
$O_{2}^{0}$ as the $\Gamma_1$ representation, finite intensities are
expected if
the primary order parameter of the ground state includes the component
$O_{2}^{0}$
among quadrupole operators.
As explained in \S \ref{sect.4.2}, the ground states
under no applied field really include the $O_{2}^{0}$ as the primary order
parameter
if we take into account all the three $S$-domains.
Because $I_{\sigma-\sigma'}(711)$ is very close
to the limiting curve for $\textbf{G} \parallel (100)$ as seen from
Fig. \ref{fig.2}, $I_{\sigma-\sigma'}(711)$ (with $h=3$)
may already exhibit the property of the limiting curve.
On the other hand, the primary order parameter of the
ground states of CeB$_6$ in the absence of the
applied field consists of the linear combination of the $\Gamma_5$-type
components. Since they do not involve $O_{2}^{0}$, no intensity
is expected in the limit of $h \rightarrow \infty$.
Thus the larger the value of $h$ is, the smaller the intensity becomes,
which explains why the intensity at $\textbf{G}=(711)$ is extremely
small in the ground state of $\Gamma_5$-type order parameter.

There is one remark on the discussion in the previous paragraph.
If our justification of why
$I_{\sigma-\sigma'}(711)/I_{\sigma-\sigma'}(533)$
is so tiny in CeB$_6$ could be
correct, we wonder why the same is not true for
$I_{\sigma-\sigma'}(511)/I_{\sigma-\sigma'}(533)$
whose experimental value is $\sim 1.22$.
In our calculation, the ratio
$I_{\sigma-\sigma'}(511)/I_{\sigma-\sigma'}(533)$ is
one order of magnitude smaller than that reported by the experiment
as mentioned before.
Consequently, our estimate gives
$I_{\sigma-\sigma'}(511)/I_{\sigma-\sigma'}(533) \sim 0.1$,
which is consistent with our justification.
We cannot find out the reason why the calculated value
$I_{\sigma-\sigma'}(511)/I_{\sigma-\sigma'}(533)$ differs
about an order of magnitude from the experimental one in CeB$_6$.
Since the ratio
$I_{\sigma-\sigma'}^{\textrm{exp.}}(511)/I_{\sigma-\sigma'}^{\textrm{exp.}}(533)
\simeq 1.22$ is inferred from the one in the $\pi-\pi'$ channel,
we wait for a direct measurement in the $\sigma-\sigma'$ channel,
however, this issue is beyond the scope of the present work.

\section{Concluding Remarks}
Owing to the extensive efforts to clarify the nature of the AFQ
phase expected from TmTe below $T_{\textrm{Q}}$, the knowledge
on the magnetic phase diagram has been
established.\cite{Matsumura1998,Clementyev1997,Link1998,
Mignot2002,Shiina1999.1,Yamamoto2007}
However, detailed understandings of ordered phase,
such as the component of the order parameter,
the nature of the microscopic multipolar interactions,
and so on,
are still rudimentary, which should be addressed.
As an attempt toward such direction, in this work,
we have investigated
the intensity of Thomson scattering from TmTe in the AFQ phase.
We have introduced a theoretical framework to
investigate the scattering amplitude
on the basis of the Rayleigh expansion of the exponential
part.\cite{Nagao2003,
Kono2004,Lovesey2002,Amara1998}
By taking the group theoretical idea into account,
we classify the selection rules determined by the orientation
of the scattering vector $\textbf{G}$.

In the absence of the external field, combining the rules and
the domain consideration,
we have obtained some qualitative criteria of the intensity
for the orientation of $\textbf{G}$, which determine the absence and/or
presence of the intensity, the degeneracy for several $\textbf{G}$
orientations,
and so on.
When we evaluate the actual intensity, however, we need information
on the ground states of the system.
We have utilized the states deduced from the $J=\frac{7}{2}$ multiplet
model developed in our previous work.\cite{Shiina2008}
For ground state with $O_{2}^{2}$ being the primary order parameter
under no external field, we have checked the three-fold degeneracy
of the intensities $I(2h+1,1,1)=I(1,2h+1,1)=I(1,1,2h+1)$
and $I(2h+1,2h+1,1)=I(1,2h+1,2h+1)=I(2h+1,1,2h+1)$,
while $I(2h+1,2h+1,2h+1)=0$.
Note that the degeneracy and the absence of intensity stated here
are concluded from the fact that the order parameter is $O_{2}^{2}$,
not from the numerical values of the expansion coefficients.

Then, we have investigated the cases in the presence of the
applied field.
The field lifts the degeneracy on the $S$-domain and breaks the cubic
symmetry of $\textbf{G}$ orientation.
For example, for $\textbf{H} \parallel (001)$, the relations
$I(2h+1,1,1)=I(1,2h+1,1) < I(1,1,2h+1)$, and
$I(1,2h+1,2h+1)=I(2h+1,1,2h+1) < I(2h+1,2h+1,1)$ hold.
For $\textbf{H} \parallel (110)$,
corresponding relations become
$I(2h+1,1,1)=I(1,2h+1,1)$ and $I(1,1,2h+1)=0$, and
$I(1,2h+1,2h+1)=I(2h+1,1,2h+1)$ and $I(2h+1,2h+1,1)=0$.
These relations have been confirmed by the numerical calculations.

Finally, we have compared our results with those obtained from another
AFQ $4f$ electron system CeB$_6$.\cite{Yakhou2001,Nagao2003}
We have concluded that the Thomson scattering intensities for TmTe at
several forbidden Bragg spots are experimentally detectable,
for instance, at $(533), (511)$, and $(711)$ in the $\sigma-\sigma'$
channel.
Then, the fact that the magnitudes of the intensities
at several spots such as $(511)$ and $(711)$ are different
significantly between for CeB$_6$ and TmTe
may be ascribed to the difference of the
components of the primary order parameters between the $\Gamma_5$-type and
$\Gamma_3$-type.
The discrimination of the component of the order parameter within the
$\Gamma_3$- or $\Gamma_5$-types
may be achieved
by the measurement of the azimuthal angle dependence
of the RXS intensity.\cite{Shiina2008}
Since our investigation lacks precise numerical information on the
weight of $K$- and $S$-domains, we should be careful when comparison of our
results with the future experimental ones will be attempted.

\section*{Acknowledgement}

This work was partly supported by Grant-in-Aid
for Scientific Research (Nos. 20540308, 21540368, and 21102520) from
the Ministry of Education, Culture, Sports, Science and Technology
in Japan.

\appendix
\section{A derivation of eq. (\ref{eq.exponent.1})}
In this Appendix, we explain a brief derivation of eq.
(\ref{eq.exponent.1}).  From eqs. (\ref{eq.state.J}) and (\ref{eq.state.LS}), 
the expectation value of $\textrm{e}^{-i \textbf{G} \cdot \textbf{r}_n}$
taken by $|0 \rangle$ becomes
\begin{equation}
\sum_{n=1}^{13}\langle 0 | \textrm{e}^{-i \textbf{G} \cdot \textbf{r}_n}
| 0 \rangle = \sum_{(\ell_z,s_z)} \sum_{(\ell_z ',s_z ')}
b_{\ell_z,s_z}^{\star} b_{\ell_z ', s_z '} f_{\ell_z,s_z:\ell_z ',s_z '},
\label{eq.app.1}
\end{equation}
where
\begin{eqnarray}
b_{\ell_z,s_z} &=& \sum_{J_z} a(J_z) C(J, J_z:\ell,-\ell_z,s, -s_z),
\label{eq.app.2} \\
f_{\ell_z,s_z:\ell_z ',s_z '} &=& \sum_{n=1}^{13}
\left\langle \ell \ell_z,s s_z \right|
\textrm{e}^{-i \textbf{G} \cdot \textbf{r}_n}
\left| \ell\ell_z ',ss_z' \right\rangle.
\end{eqnarray}
We separate eq. (\ref{eq.app.1}) into the diagonal and off-diagonal parts
as follows.
\begin{equation}
\sum_{n=1}^{13}\langle 0 | \textrm{e}^{-i \textbf{G} \cdot \textbf{r}_n}
| 0 \rangle = I_d + I_{od},
\label{eq.app.4}
\end{equation}
where
\begin{eqnarray}
I_d &=&\sum_{(\ell_z,s_z)}
|b_{\ell_z,s_z}|^2 f_{\ell_z,s_z:\ell_z,s_z},\\
I_{od} &=&
\sum_{(\ell_z,s_z)} \sum_{(\ell_z ',s_z ')}[1-\delta_{\ell_z,\ell_z'}
\delta_{s_z,s_z'}]
b_{\ell_z,s_z}^{\star} b_{\ell_z ', s_z '} f_{\ell_z,s_z:\ell_z ',s_z '}.
\nonumber \\
\end{eqnarray}

First, we consider the diagonal part.
The expectation value between the Slater determinants is evaluated as,
\begin{eqnarray}
I_d
&=& \sum_{(\ell_z,s_z)} |b_{\ell_z,s_z}|^2 \sum_{n=1}^{13} \frac{1}{13}
\nonumber \\
&\times&
\sum_{(\ell_z^n,s_z^n)\neq(\ell_z,s_z)}
\langle\langle \ell_z^n,s_z^n||
\textrm{e}^{-i \textbf{G} \cdot \textbf{r}_n}
|| \ell_z^n, s_z^n \rangle\rangle.
\end{eqnarray}
Here the double-bar state $|| \ell_z^n, s_z^n \rangle\rangle$ represents
the one electron state of the coordinate $\textbf{r}_n$, not the one
hole state.
Its representation is $\langle\langle \textbf{r}| \ell_z,s_z \rangle \rangle
=R_{4f} (r) Y_{\ell,\ell_z}(\Omega) \chi_{s,s_z}$.
Since the expectation value is independent of the electron coordinate
after the integration, we can omit it. Then, the diagonal part is
rewritten as
\begin{equation}
I_d= \sum_{(\ell_z,s_z)} |b_{\ell_z,s_z}|^2
\sum_{(\ell_z',s_z')\neq(\ell_z,s_z)} \tilde{f}_{\ell_z',s_z':\ell_z',s_z'},
\end{equation}
where
\begin{equation}
\tilde{f}_{\ell_z,s_z:\ell_z ',s_z'}
=\langle\langle \ell_z,s_z||
\textrm{e}^{-i \textbf{G} \cdot \textbf{r}}
|| \ell_z ', s_z ' \rangle\rangle. \label{eq.def.tildef}
\end{equation}
Noticing that $\sum_{(\ell_z,s_z)} |b_{\ell_z,s_z}|^2=1$, we obtain
\begin{equation}
I_d =
\sum_{(\ell_z',s_z')} \tilde{f}_{\ell_z',s_z':\ell_z',s_z'}
-\sum_{(\ell_z,s_z)} |b_{\ell_z,s_z}|^2
\tilde{f}_{\ell_z,s_z:\ell_z,s_z}. \label{eq.diagonal}
\end{equation}

Similarly, after tedious but straightforward calculations,
the off-diagonal part is rewritten as
\begin{eqnarray}
I_{od}
&=& \sum_{(\ell_z,s_z)} \sum_{(\ell_z ',s_z ')}(-)^{\ell_z-\ell_z' -1}
b_{\ell_z,s_z}^{\star} b_{\ell_z ', s_z '} \tilde{f}_{\ell_z ',s_z
':\ell_z,s_z}
\nonumber \\
&+& \sum_{(\ell_z,s_z)} |b_{\ell_z,s_z}|^2
\tilde{f}_{\ell_z,s_z:\ell_z,s_z}.
\label{eq.off-diagonal}
\end{eqnarray}
Eqs. (\ref{eq.diagonal}) and (\ref{eq.off-diagonal}) are combined
into the following expression.
\begin{eqnarray}
& & \sum_{n=1}^{13}\langle 0 | \textrm{e}^{-i \textbf{G} \cdot \textbf{r}_n}
| 0 \rangle
= \sum_{(\ell_z',s_z')} \tilde{f}_{\ell_z',s_z':\ell_z',s_z'} \nonumber \\
&-&\sum_{(\ell_z,s_z)} \sum_{(\ell_z ',s_z ')}(-)^{\ell_z-\ell_z'}
b_{\ell_z,s_z}^{\star} b_{\ell_z ', s_z '} \tilde{f}_{\ell_z ',s_z
':\ell_z,s_z}
\label{eq.app.10}
\end{eqnarray}

This expression is simplified with the help of several properties of
$\tilde{f}_{\ell_z,s_z:\ell_z',s_z'}$ which we derive below.
Substituting eqs. (\ref{eq.Rayleigh}), (\ref{eq.Gaunt}), and (\ref{eq.jGr})
into eq. (\ref{eq.def.tildef}), we get
\begin{eqnarray}
\tilde{f}_{\ell_z,s_z:\ell_z',s_z'} &=& \delta_{s_z,s_z'}
\sqrt{4 \pi} \sum_{k=0}^{\infty} (-i)^k \sqrt{2k+1}\langle j_k(G)\rangle
\nonumber \\
&\times& Y_{k,\ell_z-\ell_z'}(\Omega_G) c^k(\ell\ell_z,\ell \ell_z ').
\label{eq.flsls}
\end{eqnarray}
For a diagonal part, the summations over $\ell_z$ and $s_z$ give
\begin{eqnarray}
\sum_{(\ell_z,s_z)} \tilde{f}_{\ell_z,s_z:\ell_z,s_z} &=& 2
\sqrt{4 \pi} \sum_{k=0}^{\infty} (-i)^k \sqrt{2k+1}\langle j_k(G)\rangle
\nonumber \\
&\times& Y_{k,0}(\Omega_G)
\sum_{\ell_z=-\ell}^{\ell}c^k(\ell\ell_z,\ell \ell_z).
\end{eqnarray}
Since $\sum_{\ell_z=-\ell}^{\ell}c^k(\ell\ell_z,\ell \ell_z)=(2\ell+1)
\delta_{k,0}$
holds, we obtain
\begin{equation}
\sum_{(\ell_z,s_z)}
\tilde{f}_{\ell_z,s_z:\ell_z',s_z'} = 2 (2\ell+1)
\sqrt{4 \pi} \langle j_0(G)\rangle
Y_{0,0} = 14 \langle j_0(G) \rangle,
\end{equation}
for $\ell=3$. Now, we have an expression
\begin{eqnarray}
& & \sum_{n=1}^{13}\langle 0 | \textrm{e}^{-i \textbf{G} \cdot \textbf{r}_n}
| 0 \rangle
= 14 \langle j_0(G) \rangle \nonumber \\
&-&\sum_{(\ell_z,s_z)} \sum_{(\ell_z ',s_z ')}(-)^{\ell_z-\ell_z'}
b_{\ell_z,s_z}^{\star} b_{\ell_z ', s_z '} \tilde{f}_{\ell_z ',s_z
':\ell_z,s_z}
\label{eq.app.14}
\end{eqnarray}
in place of eq. (\ref{eq.app.10}).

A close look at eq (\ref{eq.flsls}) leads to a relation
\begin{equation}
\tilde{f}_{\ell_z,s_z:\ell_z',s_z'} =(-)^{\ell_z-\ell_z '}
\tilde{f}_{-\ell_z',-s_z':-\ell_z,-s_z}.
\end{equation}
Utilizing this yields
the second term in eq. (\ref{eq.app.14}) as
\begin{eqnarray}
&-&\sum_{(\ell_z,s_z)} \sum_{(\ell_z ',s_z ')}(-)^{\ell_z-\ell_z'}
b_{\ell_z,s_z}^{\star} b_{\ell_z ', s_z '} \tilde{f}_{\ell_z ',s_z
':\ell_z,s_z}
\nonumber \\
&=& -\sum_{(\ell_z,s_z)} \sum_{(\ell_z ',s_z ')}
b_{-\ell_z',-s_z'}^{\star} b_{-\ell_z, -s_z}
\tilde{f}_{\ell_z,s_z:\ell_z',s_z'}
\nonumber \\
&=& - \sum_{J_z} \sum_{J_z'} a^{\star} (J_z) a(J_z')
\sum_{(\ell_z,s_z)} \sum_{(\ell_z ',s_z ')}
\\
&\times&
C(JJ_z:\ell \ell_z,s s_z)
C(J J_z':\ell \ell_z',s s_z')\tilde{f}_{\ell_z,s_z:\ell_z',s_z'},
\nonumber
\end{eqnarray}
where use has been made of eq. (\ref{eq.app.2}).
Noticing the definition of $f_{J_z,J_z'}$ [eq. (\ref{eq.fjj})],
we verify the final expression is
nothing but the second term of eq. (\ref{eq.exponent.1}).


\begin{thebibliography}{9}

\bibitem{Santini2009}
P. Santini, S. Carreta, G. Amoretti, R. Caciuffo, N. Magnani, and
G. H. Lander:
Rev. Mod. Phys. \textbf{81} (2009) 807.

\bibitem{Kuramoto2009}
Y. Kuramoto, H. Kusunose, and A. Kiss:
J.~Phys.~Soc.~Jpn. \textbf{78} (2009) 072001.

\bibitem{Nakao2001}
H. Nakao, K. I. Magishi, Y. Wakabayashi, Y. Murakami, K. Koyama,
K. Hirota, Y. Endoh, and S. Kunii:
J.~Phys.~Soc.~Jpn. \textbf{70} (2001) 1857.

\bibitem{Yakhou2001}
F. Yakhou, V. Plakhty, H. Suzuki, S. Gavrilov, P. Burlet,
L. Paolasini, C. Vettier, and S. Kunii:
Phys. Lett. A \textbf{285} (2001) 191.

\bibitem{Tanaka1999}
Y. Tanaka, T. Inami, T. Nakamura, H. Yamauchi, H. Onodera,
K. Ohyama, and Y. Yamaguchi:
J. Phys.: Condens. Matter \textbf{11} (1999) L505.

\bibitem{Hirota2000}
K. Hirota, N. Oumi, T. Matsumura, H. Nakao, Y. Wakabayashi,
Y. Murakami, and Y. Endoh:
Phys. Rev. Lett. \textbf{84} (2000) 2706.

\bibitem{Matsumura2002}
T. Matsumura, N. Oumi, K. Hirota, H. Nakao,
Y. Murakami, Y. Wakabayashi, T. Arima, S. Ishihara, and Y. Endoh:
Phys. Rev. B \textbf{65} (2002) 094420.

\bibitem{Tanaka2005}
Y. Tanaka, K. Katsumata, S. Shimomura, and Y. Onuki:
J. Phys. Soc. Jpn. \textbf{74} (2005) 2201.

\bibitem{Staub2006}
U. Staub, Y. Tanaka, K. Katsumata, A. Kikkawa, Y. Kuramoto, and Y. Onuki:
J. Phys.: Condens. Matter \textbf{18} (2006) 11007.

\bibitem{Adachi2002}
H. Adachi, H. Kawata, M Mizukami, T. Akao, M. Sato, N. Ikeda,
Y. Tanaka, and H. Miwa:
Phys. Rev. Lett. \textbf{89} (2002) 206401.


\bibitem{Matsumura1998}T.~Matsumura, S.~Nakamura, T.~Goto, H.~Amitsuka,
K.~Matsuhira, T.~Sakakibara, and T.~Suzuki:
J.~Phys.~Soc.~Jpn. $\bf 67$ (1998) 612.

\bibitem{Clementyev1997}E.~Clementyev, R.~K\"{o}hler, M.~Braden,
J.-M.~Mignot, C.~Vettier, T.~Matsumura, and T.~Suzuki:
Physica B \textbf{230-232} (1997) 735.


\bibitem{Link1998}P.~Link, A.~Gukasov, J.-M.~Mignot, T.~Matsumura,
and T.~Suzuki: Phys.~Rev.~Lett. \textbf{80} (1998) 4779.

\bibitem{Mignot2002}J.-M.~Mignot, A.~Gukasov, C.~Yang,
P.~Link, T.~Matsumura, and T.~Suzuki:\textit{Proc. Int. Conf.
Strongly Correlated Electron with Orbital Degrees of Freedom
(ORBITAL2001)}.
J.~Phys.~Soc.~Jpn. {\bf 71} (2002) suppl., p. 39.

\bibitem{Shiina2008}
R.~Shiina and T.~Nagao:
J.~Phys.~Soc.~Jpn. {\bf 77} (2008) 124715.

\bibitem{Nagao2003}T.~Nagao and J.~Igarashi:
J.~Phys.~Soc.~Jpn. {\bf 72} (2003) 2381.

\bibitem{Nagao2009}
T. Nagao: \textit{The 8th Asian International Seminar on Atomic and
Molecular
Physics (AISAMP8)}, J. Phys.: Conf. Ser. \textbf{185} (2009) 012030.

\bibitem{Messiah1964}
A. Messiah: \textit{Quantum Mechanics} (North-Holland, Amsterdam, 1961).

\bibitem{Kono2004}
H. N. Kono, K. Kubo, and Y. Kuramoto:
J. Phys. Soc. Jpn. \textbf{73} (2004) 2948.

\bibitem{Lovesey2002}
S. W. Lovesey:
J. Phys.: Condens. Matter \textbf{14} (2002) 4415.

\bibitem{Keating1969}
D. T. Keating: Phys. Rev. \textbf{178} (1969) 732.

\bibitem{Amara1998}
M. Amara and P. Morin:
J. Phys.: Condens. Matter \textbf{10} (1998) 9875.

\bibitem{Cowan1981} R. Cowan: \textit{The Theory of Atomic Structure and
Spectra}
(University of California,Berkeley,1981).

\bibitem{Shiina1999.1}R.~Shiina, H.~Shiba, and O.~Sakai:
J.~Phys.~Soc.~Jpn. \textbf{68} (1999) 2105.

\bibitem{Lea1962}
K.~R.~Lea, M.~J.~M.~Leask and W.~P.~Wolf:
J.~Phys.~Chem.~Solids \textbf{23} (1962) 1381.

\bibitem{Shiina1999.2}R.~Shiina, H.~Shiba, and O.~Sakai:
J.~Phys.~Soc.~Jpn. \textbf{68} (1999) 2390.

\bibitem{Shiina2007}
R.~Shiina, H.~Shiba, and O.~Sakai:
J.~Phys.~Soc.~Jpn. \textbf{76} (2007) 094702.

\bibitem{Shiina1997}
R.~Shiina, H.~Shiba and P.~Thalmeier:
J.~Phys.~Soc.~Jpn. \textbf{66} (1997) 1741.

\bibitem{errata}
Note that the incorrect values for these ratios in the $\pi-\pi'$ channel
are printed in ref. \citen{Nagao2009}.
They should be read as
$I(511)/I(533) \simeq 3.91$ and $I(711)/I(533) \simeq 7.05$
in the $\pi-\pi'$ channel.

\bibitem{Yamamoto2007}
A.~Yamamoto, S.~Wada, and T.~Matsumura:
J.~Phys.~Soc.~Jpn. {\bf 76} (2007) 014707.






\end{thebibliography}
\end{document}